\documentclass{aa}
\usepackage{graphicx}
\usepackage{amsbsy}
\usepackage{amssymb}
\usepackage{natbib}
\bibpunct{(}{)}{;}{a}{}{,} 
\usepackage{txfonts}

\newcommand{\be}{\begin{equation}}
\newcommand{\ee}{\end{equation}}

\begin{document}

\title{The nature of long-GRB host galaxies from chemical abundances}
\titlerunning{The nature of Long-GRB host galaxies from chemical abundances}

\author{X.L. Fan\inst{1,2}\thanks{email to: fan@oats.inaf.it}
\and J. Yin\inst{3,1}
\and F. Matteucci\inst{1,2}}

\authorrunning{Fan et al.}

\institute{Dipartimento di Fisica, Sezione di Astronomia, Universit\`a di Trieste, via G.B. Tiepolo 11, I-34131, Trieste, Italy
\and I.N.A.F. Osservatorio Astronomico di
 Trieste, via G.B. Tiepolo 11, I-34131, Trieste, Italy
\and Key Laboratory for Research in Galaxies and Cosmology,
Shanghai Astronomical Observatory, Chinese Academy of Sciences, 80
Nandan Road, Shanghai, 200030, China}



\date{Received xxxx / Accepted xxxx}

\abstract{Gamma-ray bursts (GRBs) were the most energetic events after the Big Bang and they have been observed up to very high redshift.  Measurements of the chemical abundances are now available for the galaxies hosting such events,
 which are assumed to originate from the explosion of very powerful supernovae ( Type Ib/c), and provide the opportunity to study the nature of these host galaxies.}
{We identify the hosts of long GRBs (LGRBs) observed at both  low and high redshift to determine whether the hosts are  galaxies of the same type  at different cosmic epochs.}
{We  adopt detailed chemical evolution models for galaxies of different morphological type (ellipticals, spirals, irregulars) that follow the time evolution of the abundances of several chemical elements (H, He, $\alpha$-elements, Fe), and compare the results with the observed abundances and abundance ratios in galaxies hosting LGRBs.}
{We find that the abundances and abundance ratios predicted by models devised for typical irregular galaxies can reproduce the abundances of the hosts at both high and low redshift. We also find that the predicted Type Ib/c  supernova rate for irregulars is in good agreement with observations. Models for spirals and particularly ellipticals  fit neither the high-redshift hosts of LGRBs (DLA systems) nor the low redshift hosts: in particular, ellipticals cannot possibly be the hosts of gamma-ray bursts at low redshift since they exhibit little star formation, and hence no supernovae Ib/c.}
{We conclude that the observed abundance and abundance ratios in LGRBs hosts suggest that these hosts are irregular galaxies at both  high and low redshift, thus demonstra
that the host galaxies  are the same type of galaxies observed at different ages. }


\maketitle


\section{Introduction}
Gamma-ray bursts (GRBs) were the most energetic events after the Big Bang. They can be detected back to the onset of
reionization \cite[e.g.,][]{Greiner2009ApJ...693.1610G,Salvaterra2009Natur.461.1258S,Tanvir2009Natur.461.1254T} due to their brightness in the first few hours after the explosion \citep{Lamb2000ApJ...536....1L}.  Long GRBs (LGRBs) ($T_{90}\gtrsim  2$) are transient sources followed
by long-lasting afterglows emitting energy intensely across the
full range of the electromagnetic spectrum. They  are effective and informative probes of many aspects of astrophysics, such as cosmic star-formation rate (SFR) \citep{Kistler2009ApJ...705L.104K}, cosmic dust \citep{Liang2009ApJ...690L..56L}, galaxy properties \citep{Castro2008arXiv0803.2235C},  and interstellar medium (ISM) properties \citep{Levesque2010}. Whether the properties of LGRB progenitors depend or not on  metallicity is a key issue in understanding the observations. Low metallicity is indicative of a  less evolved galaxy, usually of low stellar mass, high gas-to-stellar mass ratio, and low luminosity. LGRB progenitors are understood to be low-metallicity  massive core-collapse stars \citep{Woosley1993ApJ...405..273W,Woosley2006ApJ...637..914W,Yoon2008IAUS..250..231Y} observed in association  with supernovae (SNe) Ib/c, which in turn originate in massive stars and therefore  reside in active star-forming regions \citep[e.g.][]{Galama1998Natur.395..670G,Hjorth2003Natur.423..847H,Pian2006Natur.442.1011P,Woosley2006AIPC..836..398W}. However, this low-metallicity progenitor channel is challenged  by so-called ``dark" LGRB (LGRB051022 and LGRB020819) associated with a high-metallicity environment \citep{GrahamLGRB051022,LevesqueGRB020819}.

Different systems are selected by different methods from high to low redshift. Two kinds of LGRB associated systems are selected: LGRB host galaxies  at low redshift and LGRB-associated damped Lyman-alpha systems (LGRB-DLAs)  at high  redshift.   LGRB host galaxies have been  studied by several authors using observational data \citep[e.g.][among others]{Chary2007ApJ...671..272C,Michalowski2008ApJ...672..817M,Castro2008arXiv0803.2235C,
Levesque2010,Savaglio2009,Wolf2007MNRAS.375.1049W} and simulations \cite[e.g.][among others]{Mao2010ApJ...717..140M,Chisari2010arXiv1005.4036C,Campisi2009MNRAS.400.1613C,Calura2009ApJ...693.1236C,Lapi2008MNRAS.386..608L,Nuza2007MNRAS.375..665N}. All these results find that at low redshift ($z<2$), the typical hosts of LGRB are small, star-forming, low-metallicity irregular galaxies.

At high redshift ($z >2$), as for QSO-DLA systems (log N$_{H_I}>20$ cm$^{-2}$), the  LGRB-DLAs are observed \citep[e.g.][]{Jakobsson2006A&A...460L..13J,Prochaska2007Apjs,Fynbo2009ApJS..185..526F}.
The nature of DLAs remains a matter of debate. The DLAs in QSO spectra are assumed to form in the ISM of galaxies  located in front of the QSO. The high luminosity of the QSOs usually makes it difficult detect the DLA galaxies directly. The hypothesis that these DLAs are the progenitors of present disk galaxies \citep{Wolfe1986ApJS...61..249W,Naab2006MNRAS.366..899N}  is challenged by the low metallicity ( $\simeq 1/100 Z_{\odot}$) of DLAs
and the flat age-metallicity relation of disk stars (the disk is rapidly enriched  to $\simeq 1/3 Z_{\odot}$) \citep{Akerman2004A&A...414..931A,Pettini2006fdg..conf..319P}. Using detailed chemical evolution models, \cite{Calura2003MNRAS.340...59C,Calura2009ApJ...693.1236C} investigated the nature of DLAs and  suggested that the  majority of them, including the LGRB-DLAs, may be either  spiral disks observed at large galactocentric distances,  irregular galaxies such as the LMC, or  starburst dwarf irregulars observed at different times after the last burst of star formation. On the other hand, massive elliptical galaxies are unlikely to be DLA systems even at very high redshift owing to their intensive quick enrichment, which produces high [$\alpha$/Fe] ratios, that is inconsistent with the data observed for DLAs. However, among the identified high redshift DLA host galaxies, one DLA host galaxy is a luminous Lyman-break galaxy \citep[DLA 2206-19A,][]{Moller2002ApJ...574...51M}. By comparing the luminosity  functions of DLAs and LBGs,  \cite{Wolfe2005pgqa.conf..148W} concluded that there is a significant overlap between  the DLA and LBG populations. \cite{Fynbo2008ApJ...683..321F} compared the metallicity distributions of LGRB-DLAs and LBGs. Their results support the hypothesis that LGRB-DLAs could arise from a population of LBGs that are not heavily obscured. LBGs are understood to be small young star-forming ellipticals \citep{Matteucci2002lbg,Pipino2010inpress}. Additionally, \cite{Zwaan2005MNRAS.364.1467Z} showed that in the local universe the luminosity distribution  of galaxies producing DLAs is nearly flat from $M_{B}\approx-20$ to $M_B\approx-15$, which implies that the DLA host galaxies should belong to a complex population (a single type population could not span such a wide luminosity range).
it is normally unknown in which region of the galaxy the line of sight from the QSO and LGRBs  intersect the ISM of the galaxy.  \cite{Prochaska2007ApJ...666..267P} argued that LGRB-DLAs preferentially probes denser, more depleted, higher metallicity gas located in the inner few kpc of the ISM compared to QSO-DLAs. This  idea is supported by the distribution of H{\sc i} column densities for GRB absorbers produced using a high resolution simulation of galaxy formation \citep{Pontzen2010}. In principle, it should be possible for high-redshift LGRB-DLAs to be
a phase of low-mass star-forming ellipticals, which is what we test in this paper.

In this article, we use  the most recently observed abundances of the LGRB-associated systems, LGRB-DLAs at high redshift and LGRB host galaxies at low redshift, to try to determine: i) which galaxies are the hosts of LGRBs? ii) whether the LGRB-associated systems represent an evolutionary sequence, i.e., whether the same objects at low and high redshift are just seen at different phases of their evolution? To do this,  we  adopt updated chemical evolution models for  galaxies of different morphological type that reproduce the properties of galaxies (\cite{Pipino2010inpress} for ellipticals, \cite{yin2010inpress} for irregulars and spirals), and compare our predicted chemical abundances with the observed ones.

The paper is organized as follows: in Sect. \ref{gf}, we describe our  galaxy formation scenario; in Sect. \ref{data} we summarize the observational  constraints; our results and discussions are presented in Sect. \ref{results}; and our summary and conclusions are drawn in Sect. \ref{summary}. Throughout the paper, we adopt a (0.7, 0.3, 0.7) cosmology.
\section{Galaxy formation scenario}\label{gf}
We now summarize our adopted galaxy formation models for galaxies of different morphological type.
We direct the reader to  \cite{Pipino2010inpress}  (ellipticals) and \cite{yin2010inpress} (irregular and spirals) for equations and related details. Galaxies of different morphological type have different star formation histories \citep{Matteucci2001ASSL..253.....M}. Different star formation histories produce different abundance ratios, particularly in terms of  [$\alpha$/Fe] as a function of [Fe/H], while other properties may not show such large differences, such as metallicity and stellar mass at high and low redshift, respectively (see below).

\subsection{Ellipticals}
The elliptical galaxies are  assumed to evolve following an  instantaneous mixing of gas but not according to the instantaneous recycling approximation, i.e.,  we take into account  stellar lifetimes. The model we adopt is similar to that described in \cite{Pipino2010inpress} except that we do not consider the dust production and evolution since here we compare our results only with extinction-corrected data \citep{Savaglio2006NJPh....8..195S,Savaglio2009,Levesque2010}, assuming that their dereddenings are reliable \cite[but see][]{Li2008ApJ...685.1046L}. The initial conditions  for  ellipticals allow formation by either collapse of a gas cloud  into the potential well of a dark matter halo or, more realistically, the merging of several gas clouds. In any case, the timescale for both processes, $\tau$, should be short ($<0.5$ Gyr), so that the ellipticals form in a short time.
The rapid collapse triggers an intense and rapid star-formation process,
which can be considered as a starburst that lasts until a galactic wind, powered by the thermal energy injected by stellar winds and  SN (Ia, Ibc, II) explosions, occurs. At that time, the thermal energy is equivalent to the binding energy of
gas, and all the residual gas is assumed to be lost. Numerical simulations have demonstrated that a reliable assumption is that no more than 20\% of the total blast-wave energy of supernovae should be  used to thermalize the gas \citep{Pipino2004MNRAS.347..968P}. The galactic wind develops outside-in, and after the wind the star formation stops. Therefore, in this picture the outer regions of ellipticals cease the star formation before the inner regions.
After that time, the galaxies evolve passively. An important assumption is that the efficiency of star formation, namely the SFR per unit mass of gas, increases with galactic mass: this scenario can explain the chemical downsizing \citep[for this definition of downsizing, see][]{Cowie1996AJ....112..839C}, namely that more massive ellipticals have larger [$\alpha$/Fe] ratios in their dominant stellar populations than smaller  ellipticals, as shown by \cite{Pipino2004MNRAS.347..968P}.
In principle, to explain the increase in the [$\alpha$/Fe] ratio one could assume a variable initial mass function (IMF) becoming flatter with increasing galactic mass, or a dry-merger scenario \citep{Bell2006ApJ...640..241B}, but in both of these cases other observations are  violated.
For instance, a top-heavy IMF is inconsistent with the mass-luminosity
relationship  \citep{1993Padovani,1993Matteucci}, while the dry-merger scenario can reproduce the downsizing in [Mg/Fe]-mass relation but not
the mass-metallicity relation and viceversa \citep[see][]{Pipino2008A&A...486..763P}.
One can certainly propose a selective metal loss (e.g. Fe should be lost more easily than Mg in the more massive galaxies) mechanism in any scenario to
fit this chemical downsizing, but this assumption has no a physical basis \cite[see][]{Matteucci1998A&A...335..855M}.
In this paper, we adopt a \cite{Salpeter1955ApJ...121..161S} one-slope IMF and the same stellar yields as in \cite{Pipino2010inpress}.
In particular, for massive stars we adopt the yields suggested by  \cite{Francois2004A&A...421..613F}, for the low and intermediate mass stars the yields of \cite{van1997A&AS..123..305V}, and for SNe Ia the yields of \cite{Iwamoto1999ApJS..125..439I} (model W7).
Finally, the assumed SFR is a \cite{schmidt1959ApJ...129..243S} law  that is linearly dependent on the gas density.

\subsection{Spirals and irregulars}
Spirals and irregulars are galaxies harboring  either recent or active
star formation activity. At variance with elliptical galaxies, a
slow and continuous star formation regime is assumed in spirals and
irregulars. In the irregular galaxies,  the efficiency of star formation is assumed to be lower than in spirals, an assumption that  has been  proven to be correct.

Following the work of \cite{Yin2010arXiv1005.3500Y, yin2010inpress}, we build a one-zone model for irregulars,
and assume that the galaxy is built up by continuous infall of primordial
gas until a given mass is accumulated. For the spiral disks, we develop the same model as for irregulars but with higher star formation efficiency (SFE) and higher luminous mass. As
suggested by \cite{Calura2006ApJ...652..889C}, the
main properties of local galaxies of different morphological type
could be reproduced mainly by decreasing the star formation efficiency from early to late
types. Therefore, we assume for spirals and irregulars a continuous SFR with SFE lower than the one used for ellipticals ($\simeq 3-22$ Gyr$^{-1}$),  e.g
$\simeq 1$ Gyr$^{-1}$ for spirals and $\simeq 0.1 -0.05$ Gyr$^{-1}$ for irregulars.
The irregulars are assumed to assemble by infall of gas on a relatively short timescale ($\tau \sim 1$ Gyr) and the infall mass is assumed to be $5 \times 10^{9} M_{\odot}$ for all models, whereas the spiral disks are proposed  to form by means of  a slower accreting process ($\tau \sim 7 $ Gyr) and the total infall mass is $10^{11} M_{\odot}$.
 Galactic winds triggered by SNe are also considered in the same way as for ellipticals. According to the different SFE in each model, and consequently different SN feedback, the galactic wind develops at different times in different models, which then end up with different final stellar masses.

Therefore, the main difference between models for galaxies of different morphological type in is the efficiency  star formation, which is higher in more massive, early-type objects, and consequently whether a galactic wind occurs and when.

\section{LGRB data}\label{data}
 The high-redshift LGRB-DLA  data are inferred from the rest-frame ultraviolet (UV) absorption lines.  On the other hand, low-redshift LGRB host-galaxy data are measured using the  rest-frame optical emission lines of \ion{H}{ii} regions.   The absorption lines in the UV and emission lines in the optical band have little overlap in terms of wavelength, making the detection of LGRB-DLAs and LGRB host galaxies for the same target particularly hard. In this section, we briefly summarize the observations used in this paper.

\subsection{Low redshift case: LGRB host galaxies}
The properties of GRB  host galaxies were studied by \cite{Savaglio2009} using the largest sample studied to date (40 LGRBs in total 46 GRBs).  Owing to the limitation of  measurements, 89$\%$ of the hosts are at $z\leq1.6$. In addition, all 14 LGRBs considered in this paper for which [O/H] data are available, are at $z<1$. Ten of them are studied with more detailed optical emission-line diagnostics by \cite{Levesque2010}.
\subsubsection{Morphology}
The typical morphology of  LGRB host galaxies remains unclear. The typical observed hosts of LGRB are small, star-forming, low-metallicity  galaxies. But since there is no reason for LGRBs to assume any particular galactic morphology, they should be, in principle, observed in galaxies of all morphological types.
However,  by comparing the  most faint observed star-forming LGRB host galaxies
and the rapid evolution of elliptical galaxies, we can deduce that
 any elliptical hosting a low-redshift ($z \leq 2$) LGRBs  should be small ($\sim 10^{10} M_{\odot}$) and  in its very early phase of formation.
 This formation scenario is consistent with the model of \cite{Pipino2010inpress}, where we suggested that the QSO hosts
 ($\sim 10^{12} M_{\odot} $) formed at $z\sim 8$ and the LBGs ($\sim 5\times 10^{10} M_{\odot} $)
 formed at $z\leq4$ implying that at  higher redshift ($z>1$) one could observe more
  massive  LGRB host elliptical galaxies. If this scenario were correct, we would observe
  a mass-redshift relation for LGRB hosts.

\subsubsection{ Star formation rates}
The SFRs are inferred from either the rest-frame UV ($\lambda=2800\AA$) continuum luminosity, H$\alpha$ flux, or other SFR calibrations (see data references for details).
The measured SFRs of the adopted samples span from 0.05 to 36.46 $M_{\odot}$ yr$^{-1}$ \citep{Savaglio2009}. Given the complex and uncertain SFR  measurements, here we adopt only  the critical star-forming criterion, namely that SFR$>0$, for the LGRB host galaxies.

\subsubsection{Stellar mass}
The stellar mass can be estimated from observed multi-band optical- near-infrared (NIR) photometry by performing  spectral energy distribution (SED)  fitting  \footnote{  The mass fitting is robust \citep{Shapley2005ApJ...626..698S}, despite  the metallicity, dust extinction, and age being degenerate \cite[see][who compared six evolutionary population synthesis (EPS) models and concluded that different EPS models are necessary to reproduce different stellar populations]{Chen2010A&A...515A.101C}.}{\citep{Savaglio2009}.
Some authors \citep[e.g.][]{Castro2008arXiv0803.2235C,Svensson2010MNRAS.tmp..479S} derived the stellar mass only using the rest-frame K-band flux density assuming that it is a reliable estimator.
Alternatively, the stellar mass can be estimated using an empirical relation, when other parameters
are available \citep{Savaglio2006NJPh....8..195S}.
The \cite{Savaglio2009} results indicate that the stellar mass of LGRB host galaxies are small (maximum $M_{\star}<10^{11}M_{\odot}$).
Because of the faintness of LGRB host galaxies, and the average observed stellar mass ($<M_{\star}>=10^{9.3}M_{\odot}$) lower than field galaxies \citep{Savaglio2009},  we adopt $5\times 10^9 M_{\odot}$ as the typical infall mass of our fiducial model for irregulars. On the other hand, we consider low mass elliptical galaxies and adopt  $10^{10} M_{\odot}$ as the typical mass of our fiducial model of low-mass ellipticals.
\subsubsection{Metallicity}
The metallicity (expressed as log(O/H)+12) was inferred from extinction-corrected emission-line fluxes using R$_{23}$, O3N2-metallicity relation, and auroral [O {\sc iii}] $\lambda$4363 diagnostics  by \cite{Savaglio2009} and \cite{Levesque2010}. The majority of the measured oxygen abundances are subsolar.
However, two  super-solar host galaxies are reported \citep{LevesqueGRB020819,GrahamLGRB051022}. Those super-solar metallicities challenge the assumption of a low metallicity threshold for LGRB progenitors. As we later show, their hosts may be described  by our spiral/irregular  model with SFE $\sim 0.1$ (see Fig. \ref{irsfe0.1}). In this case, the hosts may be the disks of spirals.

\subsection{High redshift case: LGRB-DLAs}
 The high-redshift GRB-DLA data are measured by the rest-frame UV absorption lines detected in the optical afterglow spectra.
Using the rest-frame UV absorption-line column densities, one  measures the cold gas-phase (T $\leq$ 1000 K)
abundance, which is  valid if we assume  negligible ionization corrections and that
the dust extinction corrections were appropriate .
The observed  abundances (e.g. Si, S, Fe)  in LGRB-DLA are subsolar, spanning  from 1/100  to near solar \citep{Savaglio2006NJPh....8..195S,Prochaska2007ApJ...666..267P,Savaglio2009}. Since \cite{Prochaska2007ApJ...666..267P} did not take into account the dust-depletion correction, we adopted the data in \cite{Savaglio2006NJPh....8..195S}.

In the literature \citep[e.g.][]{Savaglio2006NJPh....8..195S}, by adopting one reliable element abundance tracer and available dust-depletion corrections, the metallicity is generally given by that particular element abundance.
Since the elements are produced by different stars with different lifetimes, the metallicity estimated by those elements cannot  represent the true metallicity ( which is normally  estimated  by [O/H]). Measuring LGRB-DLA metallicity is a challenge at $z<1.6$, since the Ly$\alpha$ absorption line is  in the UV.  \cite{GRB090926A} presented  an extremely metal-poor LGRB-DLA system (Z $\sim$ 1/300 Z$_{\odot}$). They measured several chemical abundances and suggested that the apparent conflict  between the high [Si/Fe] and low [O/Fe] ratios found in this object is caused by the  underestimate of  the O column density.

\section{Results and discussion}\label{results}

The nature of DLAs remains unclear, but many authors have suggested that their chemical properties and star-formation histories can be reproduced by models of irregular galaxies that slowly evolve due  to  a mild SFR \citep{Bradamante1998A&A...337..338B,Romano2006MNRAS.365..759R,yin2010inpress}.
Typical elliptical galaxies are instead very unlikely to host the DLAs observed at high redshift primarily because ellipticals form rapidly \citep[in less than 1.5 Gyr, see][]{Pipino2004MNRAS.347..968P,Pipino2010inpress} consuming and then losing any residual gas so quickly that additional star formation is inhibited. Therefore, to observe an elliptical  when it is still forming stars, one should observe  it at very high redshift, higher than the typical redshift of DLAs.  Moreover, if the DLAs represent an early phase of  galaxy formation, regardless of their morphological type, we should observe DLAs with different properties ( e.g. different [$\alpha$/Fe] versus [Fe/H] relations),  which reflect the different morphological galaxy evolution history, and this is not the case.

To verify what has been stated above, we computed chemical evolution models for galaxies of different morphological type.
In Table 1, we summarize some model results: in the first column we present the model identification,  in the second column the total infall mass, in the third column  the assumed SFE, in the fourth column the predicted present-time gas mass, in the fifth column  the predicted present-time stellar mass, and in the sixth column the predicted present-time (13.7 Gyr) SFR expressed in $M_{\odot}$yr$^{-1}$. Models from 1 to 4 refer to dwarf irregulars and all assume an infall mass of $5\times 10^{9}M_{\odot}$, whereas model 5 refers to an infall mass of $10^{11}M_{\odot}$ and the disk of a spiral. Model 6 represents a small elliptical with infall mass $\sim 10^{10} M_{\odot}$ and model 7 a massive elliptical with infall mass of $10^{12}M_{\odot}$.
We note that the predicted SFRs for dwarf irregulars agree with the lowest values
estimated by  \citet{Savaglio2009}, while the SFR for the spiral disk  agrees with the highest observed values, and the gas masses  agree with the values measured for dwarf irregulars. No star formation occurs in the elliptical model after the galactic wind (occurring before 1.5 Gyr since the beginning of star formation), which implies that ellipticals cannot be the observed nearby (z$<$1) host galaxies of LGRBs, unless they just formed. On the other hand, the outermost regions of galactic disks could be associated with DLAs simply because their chemical and gaseous properties resemble those of irregular galaxies, as shown in previous works.
\begin{table*}
\begin{center}
\caption{Model predictions at 13 Gyr}
\begin{tabular}{cccccccc}
\hline
\hline
  Mod & infall mass $(M_{\odot})$&  SFE $(Gyr{-1})$ & Gas mass ($10^8M_{\odot}$) &  Stellar mass ($10^9M_{\odot})$  & SFR($M_{\odot}$yr$^{-1}$) \\
\hline
  \vspace*{-2.5mm} \\
1 & 5*$10^{9}$ & 1 & 0.079&  2.65 & 0.0079 \\
  \vspace*{-2.5mm} \\
2 &5*$10^{9}$ &0.1& 0.827& 1.34 & 0.0083\\
  \vspace*{-2.5mm} \\
3 &5*$10^{9}$ &0.05& 7.34&1.12 &0.037\\
  \vspace*{-2.5mm} \\
4 & 5*$10^{9}$&0.01 & 46.25& 0.30 &0.046 \\
  \vspace*{-2.5mm} \\
5 &$10^{11}$ &1 &31.25 &68.5   &31   \\
  \vspace*{-2.5mm} \\
6& $10^{10}$&3 &    0.0 & 4.09  & 0 \\
 \vspace*{-2.5mm} \\
7& $10^{12}$& 22& 0.0&   530 &   0\\
  \vspace*{-2.5mm} \\
\hline
\end{tabular}
\label{tab:sncolor}
\end{center}
\end{table*}

\subsection{Metallicity-redshift relation for LGRB-associated systems}\label{host_result}
To establish which galaxies  host the LGRBs at both  low and high redshift, we compare our model results with the observed abundances in LGRB-associated systems (see Figs. \ref{irsfe1}, \ref{irsfe0.1}, \ref{irsfe0.05}, and \ref{irsfe0.01}). The model with SFE=1.0 simulates the evolution of a small disk of either a spiral galaxy or  an irregular galaxy undergoing more intense star formation than the average, whereas the SFE=0.1-0.05 is typical of dwarf irregular galaxies \citep[see][]{Romano2000ApJ...539..235R, Lanfranchi2003MNRAS.345...71L}. From  Figs. \ref{irsfe1}, \ref{irsfe0.1}, \ref{irsfe0.05}, and \ref{irsfe0.01}, one can see that the value SFE=1.0 repoduces the LGRB-associated systems at neither high nor  low redshift, whereas the model for irregulars galaxies with SFE=0.1-0.05 can closely  fit all the data.
On the other hand, values of SFE$<$0.05 do not fit the data.
This comparison clearly indicates that the associated systems of the studied LGRBs are dwarf irregular galaxies
in different evolutionary phases. In elliptical galaxies, there is no star formation after the galactic wind, therefore they can be rejected as hosts of LGRBs at low redshift. Since  ellipticals are very rapidly enriched, by a  fine-tuning of the formation redshift parameter, one can always find a model of elliptical capable of reproducing any subsolar log(X/H)+12 data at high redshift, but this would be  an arbitrary assumption. We therefore need abundance ratios to test this situation (see Figs. \ref{si_fe} and \ref{s_fe} in the next section).

In Fig. \ref{spsfe1}, we show again the log(X/H)+12 versus redshift  plot predicted by a model with SFE=1.0 but an infall mass of $10^{11}M_{\odot}$ that should represent the disk of a spiral galaxy such as the Milky Way (model 5). In this model, we also assumed that the gas accreted more slowly  than dwarf irregulars. Detailed chemical evolution models for the Milky Way \cite[e.g.][]{Chiappini1997ApJ...477..765C,Boissier1999MNRAS.307..857B,Hou2000A&A...362..921H,Chiappini2001ApJ...554.1044C,Yin2009A&A...505..497Y} predict that the Milky Way disk formed in several Gyrs and inside-out. Here, we present a simplified one-zone model for a galactic disk that predicts average abundances, which helps to verify the behavior of the age-metallicity relation of  a more massive galaxy with a higher SFE than typical irregulars. Figure \ref{spsfe1} shows that this model, as happens for the model with infall mass $5\times 10^{9}M_{\odot}$ and SFE typical of a spiral disk (model 1), cannot fit any of the LGRB hosts and therefore should be rejected. We note that we  assumed a common redshift of galaxy formation for all the models ($z_f=10$) irrespective of the galaxy morphological type because  old stars are in general found in every galaxy.  If  galaxies formed later, clearly they can not be the hosts of high-redshift LGRBs. However, irregulars may be the hosts of low-redshift LGRBs  if they formed at redshift z=1, although in this case a slightly lower SFE would be required.

\subsection{Abundance ratios}
The abundance ratios of chemical elements, which have different timescales of production (such as $\alpha$-elements and Fe), can be used as useful criterion to study the star formation history in galaxies \cite[see][]{Matteucci2001ASSL..253.....M}.
In Figs. \ref{si_fe} and \ref{s_fe}, we compare the [Si/Fe] and [S/Fe] ratios predicted by irregular and elliptical models with the available data for  LGRB-DLAs.
The data are derived from \cite{Savaglio2006NJPh....8..195S} and \cite{GRB090926A}.
Although the error bars in the data are very large, the results imply that most of the data
cannot be well fitted by the predictions of elliptical models. They are instead reproduced by the predictions of irregular models with a SFE=$0.05$ Gyr$^{-1}$, thus confirming the results shown in Figs \ref{irsfe1}, \ref{irsfe0.1}, \ref{irsfe0.05}, and \ref{irsfe0.01}.
However, more accurate data are necessary in the future to confirm this result.

In the case both of Si and S, because of the large error bars in the observational data, the predictions
of the elliptical models are marginally acceptable.
In general, our predictions indicate that ellipticals, when forming stars ( a passive evolutionary phase is of no interest because is without star formation and SNe Ib/c  are connected to star formation since they originate in massive stars), have higher [$\alpha$/Fe] ratios than spirals and irregulars and that their evolution is much shorter. Hence, if one measure were to the  [$\alpha$/Fe] ratios of the gas in star-forming ellipticals at high redshift, one  should find super-solar values. However, ellipticals are predicted to remain in a phase where their gas has a metallicity [Fe/H] in common with those of DLAs (between -2.0 and -3.0) for a very short time interval ($<$ 0.1 Gyr), as shown in Fig. \ref{FeH_t}. Therefore, it is very difficult to observe these galaxies in this short time interval, another reason for excluding even small ellipticals from being DLA systems.
If the DLAs did represent a particular evolutionary phase of a galaxy, regardless of  morphological type,
we should then observe DLAs with a wide range of  properties (e.g. [$\alpha$/Fe]), which reflecting the different
star formation histories of galaxies of morphological type.

Since LGRBs are associated with SNe Ib/c, we  also computed these rates for different galaxies, and compared the predicted present-time SN Ib/c rate with the observed one, as shown in Fig. \ref{sn1bc}. To compute this rate, we assumed that the progenitors of these SNe are either Wolf-Rayet stars with $M_{in} > 25 M_{\odot}$ or massive binary systems in the mass range $12 M_{\odot} \le M_{in} \le 20 M_{\odot}$, as described in detail in \cite{Bissaldi2007A&A...471..585B}. For the observational  SN Ib/c rate,
we assumed the value of the SN Ib/c rate per unit mass provided by \cite{Mannucci2005A&A...433..807M} for irregular galaxies, which is $0.54 ^{+0.66}_{-0.38}$ SNe (100yr)$^{-1} 10^{-10}M_{\odot}$. We then multiplied
this value by the predicted present-time stellar mass in each model (Table 1).
The results show that our irregular models  predict SN Ib/c rates in reasonable agreement with observations.
In addition, the observed GRB redshift distribution peaks around z=1 (Fig. 1 in \cite{Savaglio2009})  in reasonable agreement with our predicted peak of SN Ib/c rate  (at z=1 corresponding to a galactic age $\sim$ 5Gyr, see Fig. 9).  The sharp break in the SN Ib/c rate for ellipticals and irregulars with SFE=0.05 occurs because at that time the galactic wind develops (SFE=0.01 is too low to develop a wind). Gas is lost through the wind and the gas surface density decreases sharply (see Sec. \ref{gf} for details). As a consequence, the SFR  drops  and the SN Ib/c  rate then  also drops.

\section{Summary}\label{summary}
We have compared the data of  LGRB-associated systems, LGRB-DLAs at high redshift, and LGRB host galaxies at low redshift, with  chemical evolution models for galaxies of different  morphological type. We have attempted to answer the following questions: i) which galaxies are the hosts of LGRBs? ii) are the LGRB systems  part of an evolutionary sequence, in other words are they the same objects at low and high redshift as seen in different phases of their evolution?

Our conclusions can be summarized as follows:
\begin{enumerate}
\item If the observed high-redshift LGRB-DLAs and local LGRB host galaxies belonged to an evolutionary sequence, they should be  irregulars with a common galaxy-formation redshift as high as $z_{f}=10$, observed at different phases of their evolution. We can fit the majority of these systems using slowly evolving irregular models with a star formation efficiency, SFE=0.1-0.05. The adopted models were previously tested on local dwarf irregulars \cite[see][]{Yin2010arXiv1005.3500Y,yin2010inpress} and they may  reproduce their  properties.

\item We tested models with SFE=1.0, which resemble the evolution of the disk of the Milky Way \citep{Chiappini1997ApJ...477..765C,Francois2004A&A...421..613F,Cescutti2007A&A...462..943C}. We considered the average properties of such a disk, and concluded that they fit neither high nor  low redshift data since they predict too high absolute abundances. We cannot exclude, however, that they correspond to the outermost regions of spiral disks, since their properties are similar to those of irregulars \cite[see][]{Matteucci1997A&A...321...45M}.

\item The rapid chemical enrichment at high redshift and subsequent passive evolution, following the occurrence of a  galactic wind  several Gyrs ago, of elliptical  galaxies,  suggests that they cannot be LGRB host galaxies at low redshift and that they are very unlikely  hosts of  LGRB-DLAs even at high redshift.  The high observed [$\alpha$/Fe] ratios in ellipticals
indicate in particular that most LGRB-DLAs can be neither a phase nor a part of ellipticals.

\item We have confirmed that the star formation history is the main driver of galaxy evolution. In particular, the properties of galaxies of different morphological type can be reproduced  by simply changing the efficiency of star formation. By comparing the observed and predicted chemical properties of LGRB hosts, we have shown that they are most closely fitted by galaxy models with a lower SFE than either spirals or ellipticals, which is typical instead of irregular galaxies.

\item  The elliptical models to which we compared the data refer to a very massive elliptical of infall mass $10^{12} M_{\odot}$, which is reproduced well by assuming a SFE=22 Gyr$^{-1}$, and to a small elliptical of infall mass $10^{10}M_{\odot}$, which is reproduced well  by assuming a lower SFE=3 Gyr$^{-1}$. These assumptions reflect the downsizing in star formation that is necessary to reproduce the majority of the chemical and photometric properties of ellipticals. \citep[e.g. downsizing in chemical enrichment, see ][]{Pipino2004MNRAS.347..968P,Pipino2006MNRAS.365.1114P,Pipino2008A&A...486..763P}.
 According to our anti-hierarchical galaxy formation scenario,  we should with future data observe a mass-redshift relation for LGRB hosts.
\end{enumerate}

\acknowledgements  Our thanks go to the anonymous referee
for valuable comments and useful suggestions. J.Y. and F.M. acknowledge the financial support from
PRIN2007 from Italian Ministry of Research, Prot. no.
2007JJC53X-001. J.Y. thanks the hospitality of the Department of
Physics of the University of Trieste where this work was
accomplished. J.Y. also thanks the financial support from the
National Science Foundation of China No.10573028, the Key Project
No.10833005, the Group Innovation Project No.10821302, and 973
program No. 2007CB815402.

\begin{figure}
\includegraphics[width=3in]{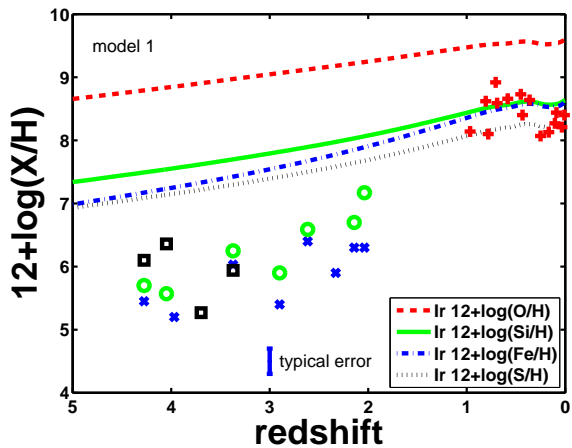}
\caption{ 12+log(X/H) as a function of redshift. Predicted lines are from the irregular galaxy model with SFE=1, as indicated in the figure. High redshift ($z>2$) data are LGRB-DLA systems and  low redshift ($z< 1$) data are LGRB host galaxies. Red pluses, green circles, blue crosses, and black squares are relative to O, Si, Fe, and S abundances, respectively. Typical error in data is shown by a vertical line.  Data are derived from \cite{Savaglio2006NJPh....8..195S}, \cite{Levesque2010}, \cite{Savaglio2009}, and \cite{GRB090926A}.}\label{irsfe1}
\end{figure}

\begin{figure}
\includegraphics[width=3in]{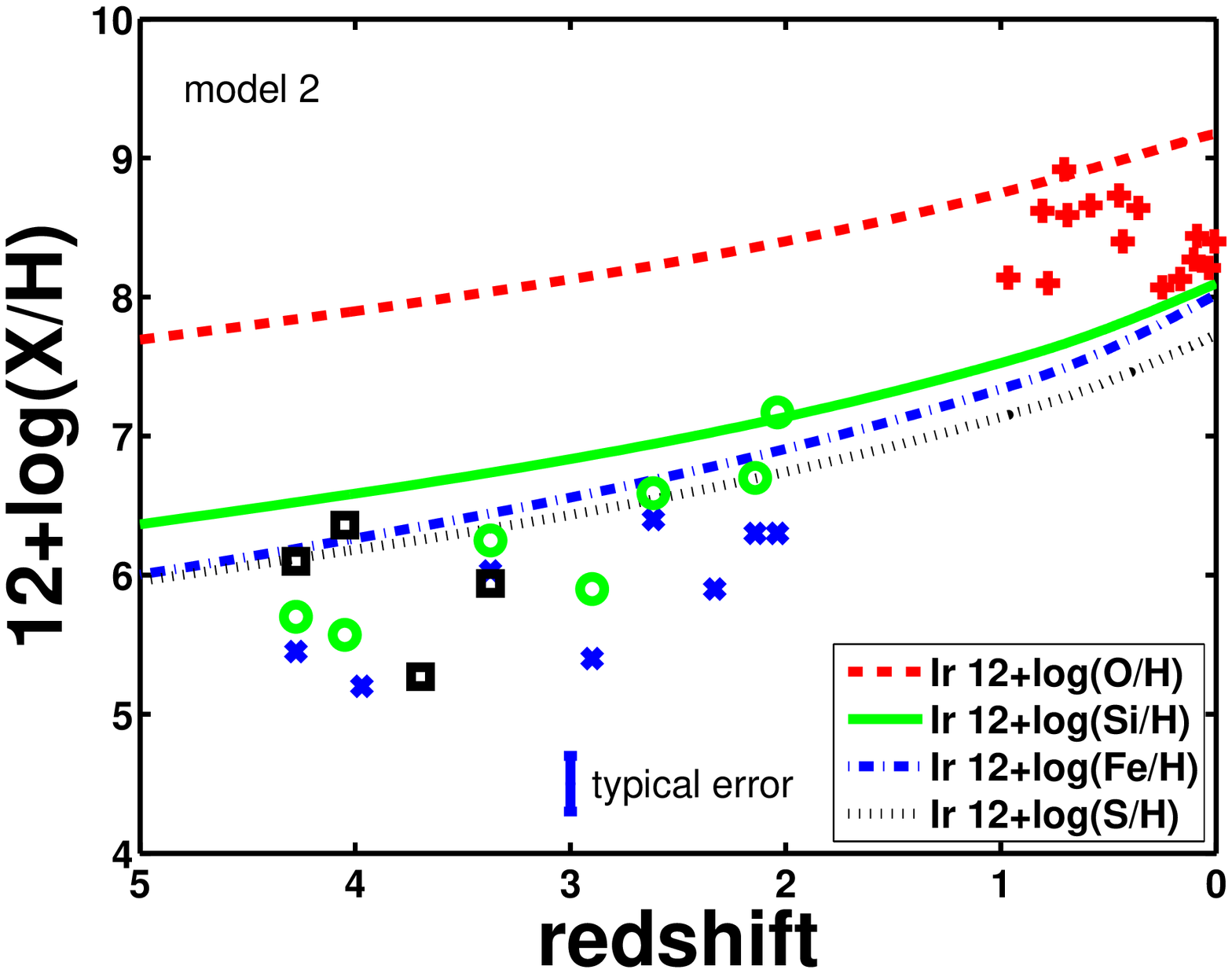}
\caption{ 12+log(X/H) as a function of redshift. The same as Fig. \ref{irsfe1} but for SFE=0.1. }\label{irsfe0.1}
\end{figure}

\begin{figure}
\includegraphics[width=3in]{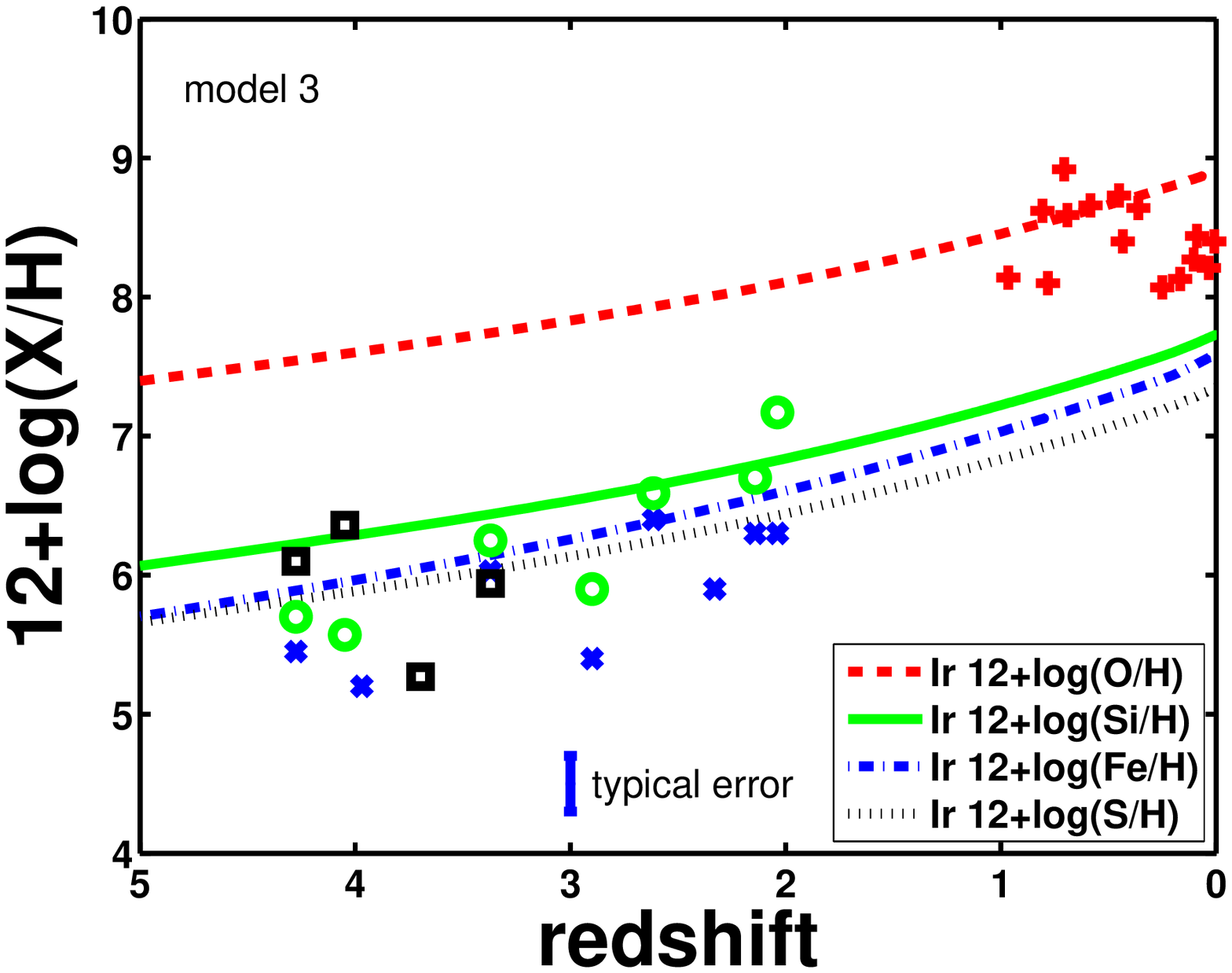}
\caption{ 12+log(X/H) as a function of redshift. The same as Fig. \ref{irsfe1} but for SFE=0.05.}\label{irsfe0.05}
\end{figure}

\begin{figure}
\includegraphics[width=3in]{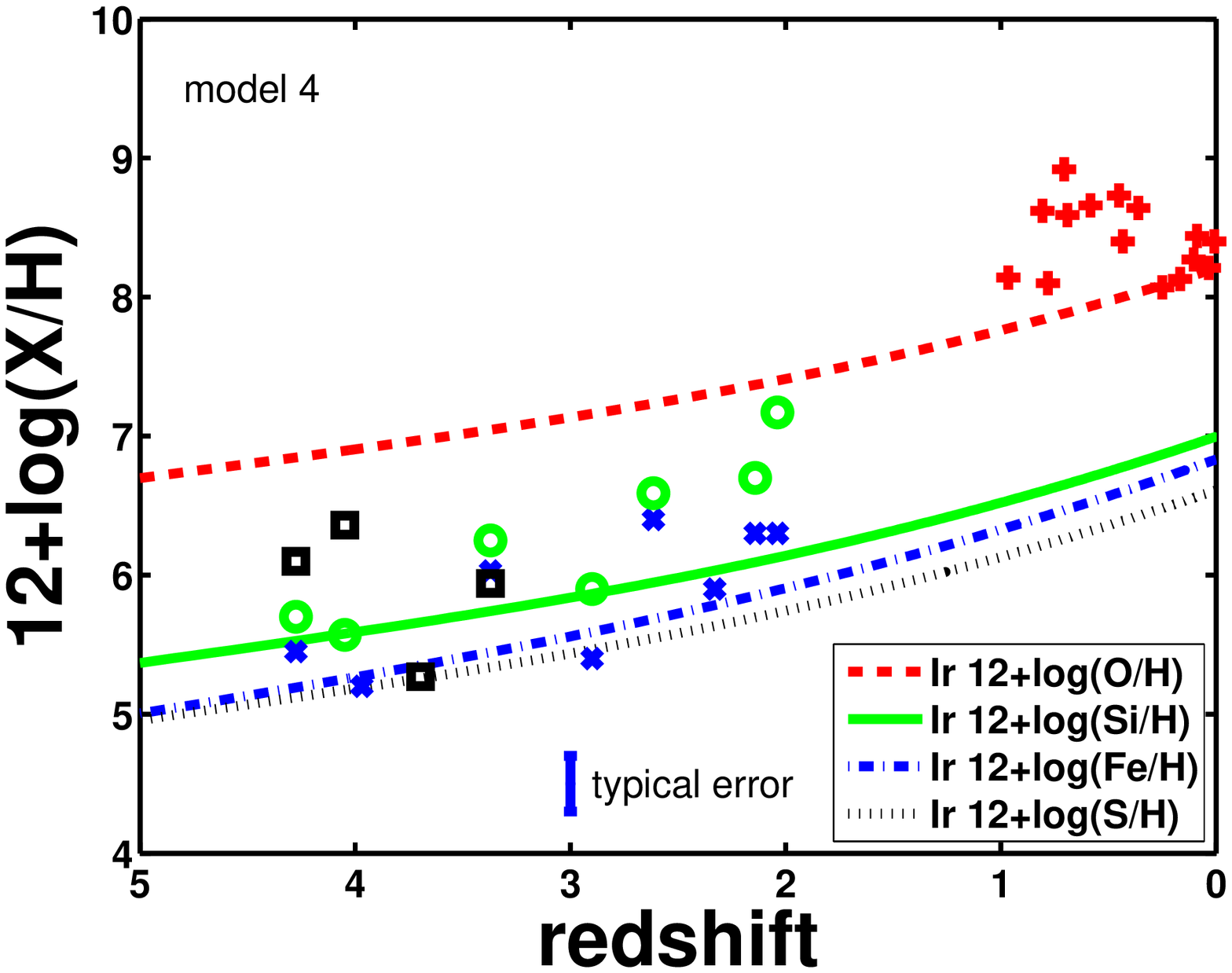}
\caption{12+log(X/H) as a function of redshift. The same as Fig. \ref{irsfe1} but for SFE=0.01. }\label{irsfe0.01}
\end{figure}

\begin{figure}
\includegraphics[width=3in]{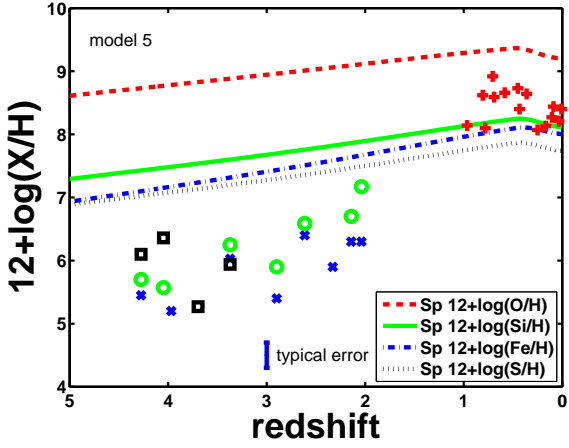}
\caption{12+log(X/H) as a function of redshift. The same as Fig. \ref{irsfe1} but for the spiral galaxy model with  SFE=1 and infall mass $10^{11} M_{\odot}$}.\label{spsfe1}
\end{figure}

\begin{figure}
\includegraphics[width=3in]{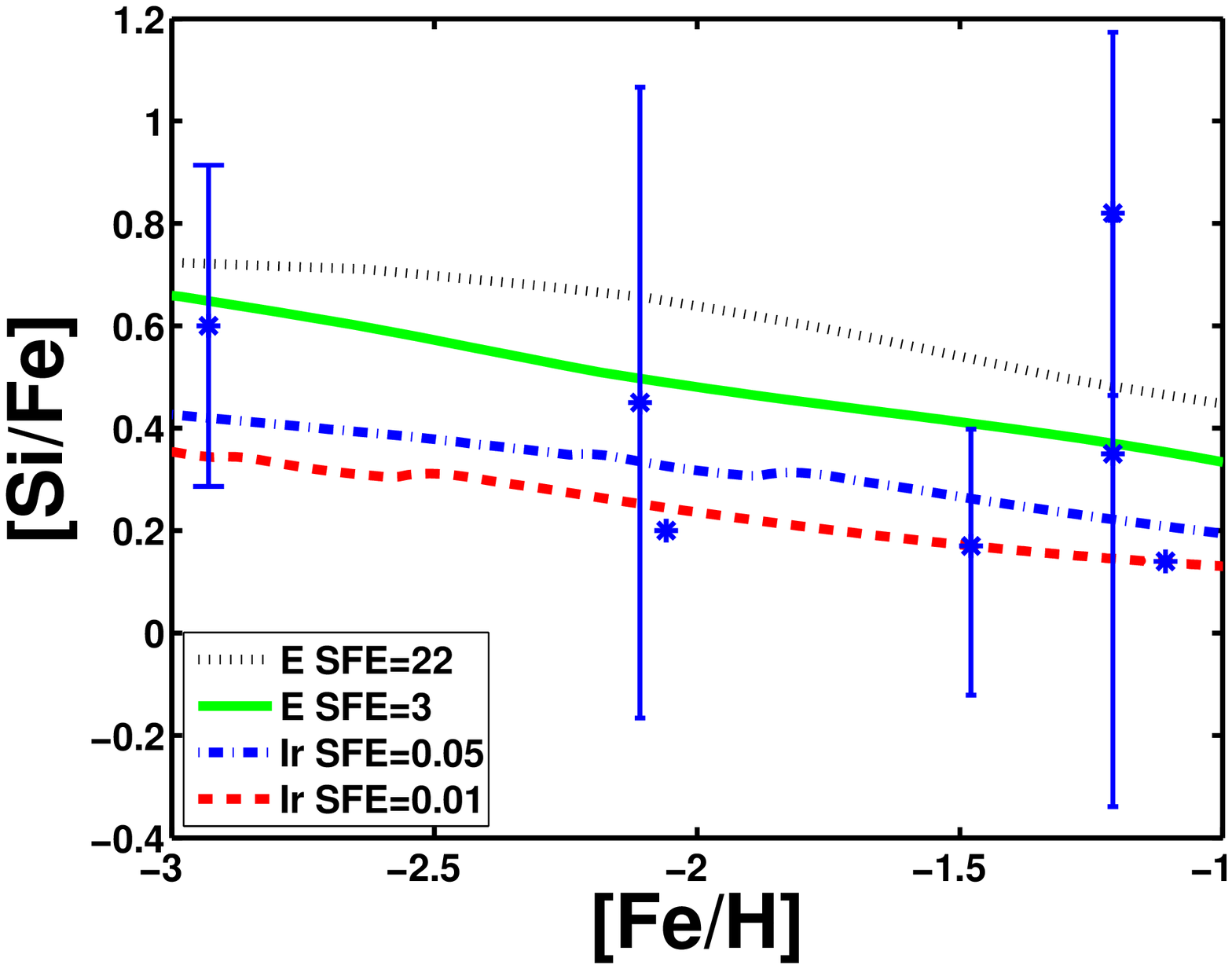}
\caption{ [Si/Fe] as function of  [Fe/H] of  LGRB-DLA systems. The predictions are  our from chemical evolution models of  elliptical galaxies with SFE=22 (black  dotted line) and SFE=3 (green solid line), and of irregular galaxies with SFE=0.05 (blue dashed-dotted line) and SFE=0.01 (red dashed line) (see text). Data are derived from \cite{Savaglio2006NJPh....8..195S} and \cite{GRB090926A}.}\label{si_fe}
\end{figure}
\begin{figure}
\includegraphics[width=3in]{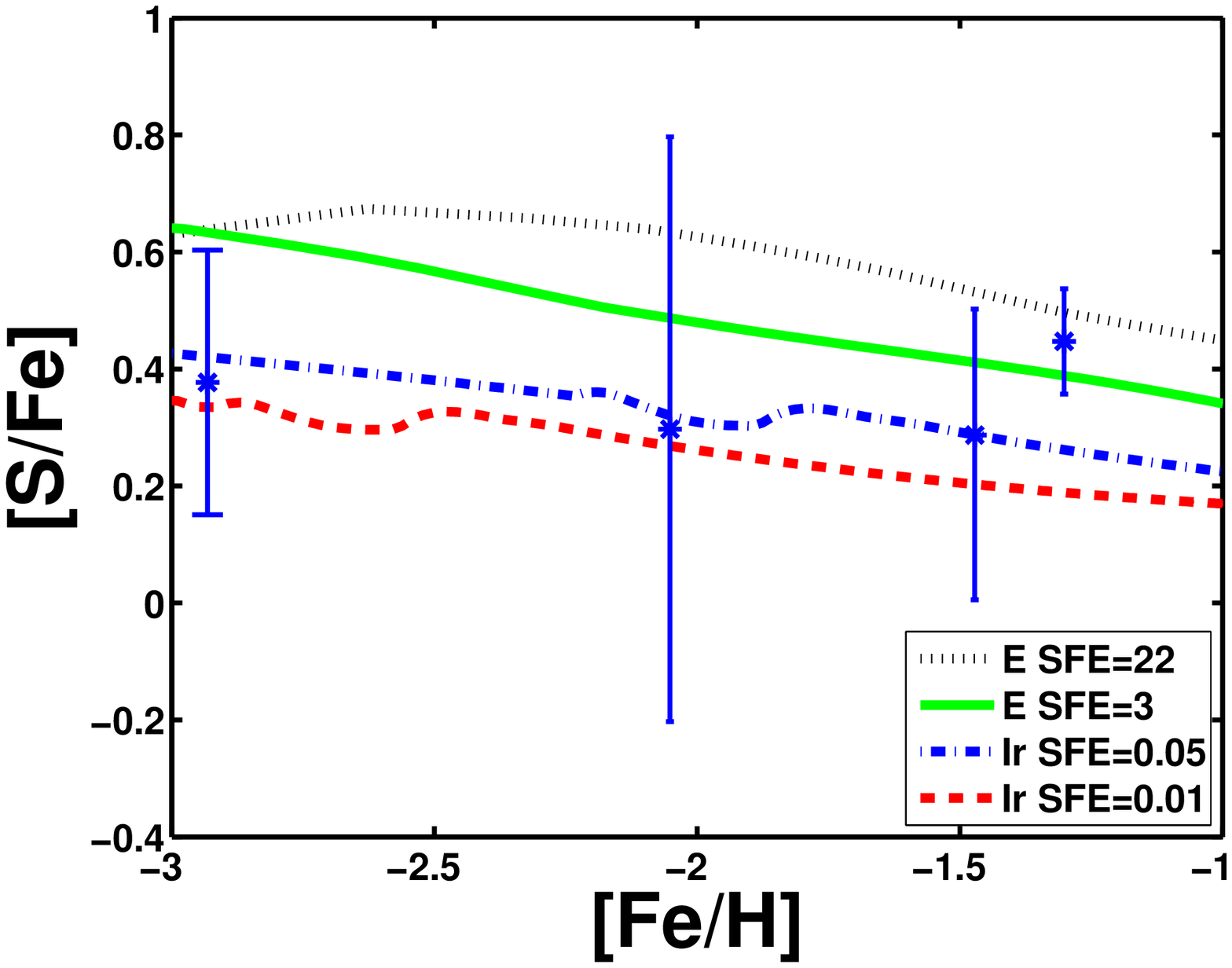}
\caption{ [S/Fe] as function of  [Fe/H] of  LGRB-DLA systems. The predictions are  from chemical evolution models for the elliptical galaxies with SFE=22 (black  dotted line) and SFE=3 (green solid line), and for irregular galaxies with SFE=0.05 (blue dashed-dotted line) and SFE=0.01 (red dashed line) (see text). Data are derived from \cite{Savaglio2006NJPh....8..195S}.}\label{s_fe}
\end{figure}

\begin{figure}
\includegraphics[width=3in]{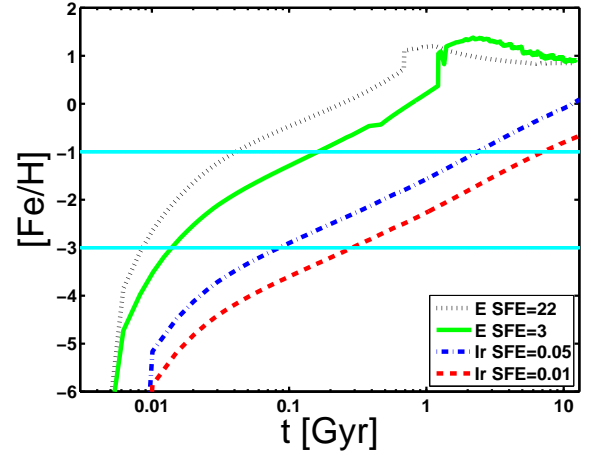}
\caption{Predicted [Fe/H] versus  galactic age for two irregular and two elliptical models, as indicated in the figure. The horizontal lines mark  [Fe/H]$=-1.0$ and [Fe/H]$=-3.0$.}
\label{FeH_t}
\end{figure}

\begin{figure}
\includegraphics[width=3in]{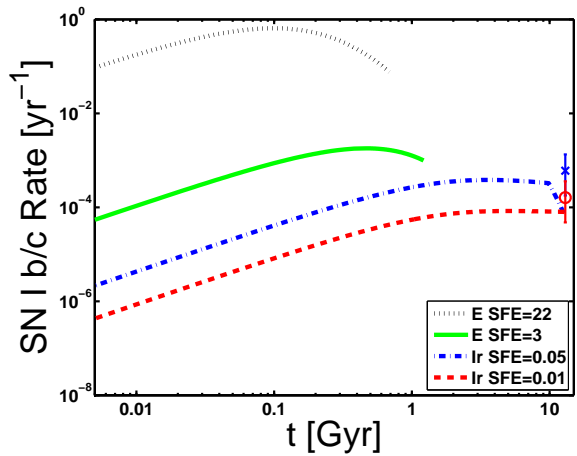}
\caption{Predicted SNIb/c rate evolution as a function of  galactic age for irregulars and ellipticals. The predictions are  from our chemical evolution model of a  elliptical galaxy  assuming  the star formation efficiency  SFE=22 (black  dotted line), SFE=3 (green solid line)  and an irregular galaxy assuming SFE=0.05 (blue  dashed-dotted line) and  SFE=0.01 (red dashed line) (see text). Blue cross and red circle are the observed  SN Ib/c rates, derived by multiplying the  SN Ib/c rate per unit mass in irregular galaxies \citep{Mannucci2005A&A...433..807M} by the predicted present stellar mass in each model. }\label{sn1bc}
\end{figure}


\bibliographystyle{aa}
\bibliography{lgrb_ref,chemical_model,SN,dust,sed_fitting,galaxy_formation,review,DLA,empirical_relation,LBG}

\end{document}